%% file: main_avec_nom.tex
\documentclass[12pt, a4paper]{article}

\usepackage[utf8]{inputenc}
\usepackage[T1]{fontenc}
\usepackage[english]{babel}
\usepackage{geometry}
\usepackage{tikz}
\usepackage{multirow}
\usepackage{makecell}
\usetikzlibrary{positioning}
\usepackage{graphicx}
\usepackage{setspace}
\usepackage{amsmath, amssymb}
\usepackage{mathtools}

\usepackage{graphicx}
\usepackage{booktabs}
\usepackage{array}
\usepackage{hyperref}
\usepackage{url}
\usepackage{natbib}
\setcitestyle{authoryear, maxcitenames=1}
\usepackage{footmisc}
\usepackage{caption}
\usepackage{float}
\usepackage{cleveref}
\usepackage{tabularx}

\title{
    \textbf{Mapping NGFS Climate Pathways to GICS Sectors: Application to Portfolio Decarbonization} \\[0.3em]
    \large  An Emissions-Based Framework for Portfolio Climate Alignment 
}
\author{
    Maxime Jourand\thanks{EDF Lab. E-mail: \texttt{maxime.jourand@edf.fr}}
    \and
    Adel Medjabri\thanks{Institut Louis Bachelier. E-mail: \texttt{adel.medjabri@institutlouisbachelier.org}}
}
\date{November 2026}

\begin{document}

\maketitle
\vspace{-1cm}
\thispagestyle{empty}

\begin{abstract}
\noindent
Portfolio climate alignment is limited by the mismatch between asset managers' sector classifications and those of climate scenarios. This article proposes an NGFS–GICS mapping based on the materiality of Scope 1 and 2 GHG emissions rather than on semantic similarity. The mapping feeds a Sectoral Decarbonization Approach that down-scales NGFS trajectories into GICS sector pathways at portfolio level, adjusted for regional exposure. Applied to two illustrative ETF indices, it shows how sectoral composition and regional trajectories shape a portfolio's financed emissions pathway. The framework is transparent, asset-class agnostic and usable for internal exercises such as climate stress tests.
\end{abstract}

\hfill

\newpage


\section{Introduction}

\input{1_Intro}

\section{GICS-NGFS mapping for GHG materiality}
\label{sec:GICS-NGFS mapping}
\input{2a_NGFS_GICS_jointure}
\input{2b_Results}
\input{2c_Discussion_conclusion}
\section{Sectoral Decarbonization Approach with GICS-NGFS mapping}
\input{3a_Materials_and_methods}
\input{3b_Application}

\input{3c_Discussions_conclusions}
\section{Conclusion}
\input{4_Conclusions}
\newpage
\section*{Acknowledgments}
\input{5_Acknowledgments}

\newpage
\section*{Declaration of interest statement}

The authors declare that they have no financial conflicts of interest related to the findings of this article.

\section*{AI disclosure}

The authors used generative AI tools exclusively for language editing and readability improvements. All analyses, interpretations, and conclusions are those of the authors.
\newpage
\nocite{*}
\bibliography{references}

\newpage
\appendix

\section{Appendix A: Identification and treatment of specific cases}

\paragraph{NGFS-1 Residential and Commercial}

The NGFS Residential and Commercial sector presents a particular challenge for the mapping exercise: Scope 1 and Scope 2 emissions associated with this sector primarily arise from building-related activities, including electricity consumption, heating, and air conditioning in office and commercial premises.

Following the principle of matching sectors according to the materiality of their Scope 1 and Scope 2 emissions, the NGFS Residential and Commercial sector was mapped to three GICS-1 sectors:

\begin{enumerate}
    \item Communication Services;
    \item Financials;
    \item Real Estate.
\end{enumerate}

For these sectors, emissions are largely driven by the operation of office buildings and commercial facilities, making them directly comparable to the emissions sources represented by the NGFS R\&C category.

However, the mapping is less straightforward for a number of additional NACE sectors. In these cases, emissions may originate both from office-related activities and from other operational processes with potentially significant GHG emissions. For example, NACE code 72 (Scientific Research and Development) may encompass activities ranging from office-based research to laboratory and testing activities associated with higher emissions levels.

Approximately 50 NACE sectors were identified as presenting this type of ambiguity. To avoid introducing potentially misleading correspondences, a conservative approach was adopted whereby these sectors were excluded from the mapping. Consequently, the NGFS Residential and Commercial sector is linked solely to the Communication Services, Financials, and Real Estate GICS-1 sectors.

\paragraph{NGFS-1 AFOLU} The NGFS AFOLU\footnote{Agriculture, Forestry, and Other Land Use} sector presents a specific challenge, as the GICS classification does not contain a dedicated GICS-1 sector exclusively covering agricultural activities. Using the NACE--NGFS correspondence table developed by the University of Zurich and IIASA, the following NGFS-2 categories were identified within the AFOLU sector:

\begin{itemize}
    \item Agricultural Production|Non-Energy|Crops;
    \item Agricultural Production|Non-Energy|Livestock;
    \item Land Cover|Forest|Managed.
\end{itemize}

A review of the GICS methodology identified three potentially relevant GICS-2 categories:

\begin{itemize}
    \item Agricultural Products (GICS-1 Consumer Staples);
    \item Packaged Foods \& Meats (GICS-1 Consumer Staples);
    \item Forest Products (GICS-1 Materials).
\end{itemize}

Because the proposed mapping focuses exclusively on Scope 1 and Scope 2 emissions, only GICS categories directly associated with agricultural and forestry production activities were retained. Consequently, the GICS-2 Agricultural Products category was mapped to the NGFS AFOLU sector, as it includes crop and livestock production activities. In contrast, the GICS-2 Packaged Foods \& Meats category was excluded because its emissions profile is primarily associated with downstream processing activities, which are more closely related to value-chain emissions and therefore fall outside the scope of the present analysis.

In addition, the GICS-2 Forest Products category was linked to the AFOLU sector, given the explicit inclusion of forestry activities in the NGFS definition. The resulting mapping therefore associates the NGFS AFOLU sector with two GICS-2 categories: Agricultural Products and Forest Products.

\paragraph{GHG emissions from the manufacturing of transportation equipment}

The proposed mapping does not distinguish between the NGFS variables \textit{Final Energy} and \textit{Energy Service}. From a corporate perspective, annual energy demand (\textit{Energy Service}) broadly corresponds to annual energy consumption (\textit{Final Energy}). Consequently, when focusing exclusively on Scope 1 and Scope 2 emissions, the energy demand associated with transportation equipment manufacturers is allocated to the NGFS-1 \textit{Industry} sector rather than to the \textit{Transportation} sector.

This choice reflects the fact that the direct emissions and purchased-energy emissions of transportation equipment manufacturers primarily arise from manufacturing activities. By contrast, emissions associated with the use of vehicles after their sale belong to the use phase of the products and are therefore captured as downstream Scope 3 emissions. Since Scope 3 emissions are excluded from the present analysis, transportation equipment manufacturing activities are mapped to the \textit{Industry} sector.

\paragraph{GHG emissions from manufacturers of metal} Regarding the NACE-2 code “Manufacture of basic metals”, it is important to note that we consider only the production of metals used in nuclear energy generation, and not the production of all metals covered by this activity. Consequently, we link it to the NGFS-2 “Other Energy Supply” sector.

\paragraph{GHG emissions from leasing and car rental companies}

Vehicle rental and leasing activities initially appeared to be natural candidates for inclusion within the NGFS \textit{Transportation} sector, as the emissions associated with the use of leased vehicles are substantially higher than those generated by conventional office activities, which are typically captured within the NGFS \textit{Residential and Commercial} sector.

However, because the present mapping focuses exclusively on Scope 1 and Scope 2 emissions, the NACE-2 \textit{Rental and Leasing Activities} category was ultimately excluded. According to the GHG Protocol, the majority of emissions associated with leasing and vehicle rental activities fall under Scope 3, notably Category 13 (\textit{Downstream leased assets}). By contrast, the direct emissions and purchased-energy emissions generated by leasing companies are generally limited relative to the emissions arising from the use of leased assets.

Consequently, including this sector within the mapping would introduce emissions that lie outside the scope of the present analysis. A conservative approach was therefore adopted, whereby leasing and vehicle rental activities were excluded in order to maintain consistency with the Scope 1 and Scope 2 focus of the methodology.

\end{document}

%% file: 1_Intro.tex
In line with the objectives of the Paris Agreement (COP21, 2015), financial institutions must integrate ESG\footnote{Environmental, Social and Governance} factors into their activities. Moreover, the regulatory landscape has evolved significantly (CSRD, SFDR), notably with the introduction of double materiality principle requiring companies to disclose both their environmental impacts and the financial risks associated with sustainability issues. In this context, portfolio managers are increasingly expected to lead climate alignment exercises, which evaluate the consistency of financial portfolios or economic activities with predefined climate pathways, such as those associated with the Paris Agreement. These approaches typically rely on forward‑looking estimates of GHG\footnote{Greenhouse gas} emissions and scenario-based benchmarks to assess whether current trajectories are compatible with long-term climate objectives.

To help portfolio managers in this exercise, there are numerous tools available. In Alignment Cookbook II \citep{ILB2024AlignmentCookbook2}, the Institut Louis Bachelier analyzed more than 50 methodologies, including the Sectoral Decarbonization Approach. This method defines sector-specific GHG trajectories consistent with long-term net-zero pathways. It is widely used across portfolio alignment methodologies as it relies on scientific climate scenarios (e.g., IPCC or IEA) and provides granular, sector-differentiated projections. However, this approach is not universal since the results depend on both sectoral classifications and scenario assumptions, which vary across data providers\footnote{ICB (Industry Classification Benchmark) from FTSE Russell, BICS (Bloomberg Industry Classification System) from Bloomberg, among others}.

The multiplicity of sectoral classifications creates a challenge for portfolio managers in alignment exercises: when relying on external datasets, the classification used by data providers often differs from the one applied internally, creating a need for translation between systems. In the absence of a consistent mapping between sectoral classifications, portfolio managers face significant difficulties in exploiting external data, ensuring consistency in alignment assessments, and producing robust portfolio-level results. 
This raises a key question: how can two sectoral classifications with fundamentally different structures and scopes be effectively linked? More specifically, how can practitioners incorporate the materiality of GHG emissions into a classification framework that remains consistent with their internal practices?

Accordingly, this paper contributes to this issue with a methodology combining two steps : 1) a practical mapping between GICS\footnote{MSCI, The Global Industry Classification Standard (https://www.msci.com/indexes/index-resources/gics)} and NGFS\footnote{Network for Greening the Financial System} sectoral classifications, with a focus on explicitly capturing GHG emissions materiality across sectors ; and 2) a Sectoral Decarbonization Approach down-scaling NGFS projections into GICS specific GHG trajectories for a given manager's portfolio.

GICS is a classification developed by S\&P and MSCI, widely used to define sectors in stock market indices. This classification is commonly adopted by asset managers as it provides a standardized framework for categorizing economic activities and enables practitioners to compare portfolio performance across sectors. The NGFS (Network for Greening the Financial System) classification is designed to support variables used with NGFS climate scenarios. It differs significantly from GICS, as it does not aim to categorize economic activities, but rather to capture differences in climate risk exposure across sectors and regions. Each sector is associated with specific variables—whether economic, physical or climate-related—reflecting its role within scenario analysis. As such, this classification is both focused and operational for climate modeling purposes.
A mapping between these classifications can be highly valuable for practitioners, supporting climate alignment assessments, forward-looking indicator analyzes, and regulatory stress testing.

Sectoral mappings related to GICS or NGFS have been studied, notably: a NACE-CPRS-IAM mapping by \citet{iam_nace} and a NACE-GICS mapping by the Banque de France (2024). In 2026, Battiston et al. propose a GICS-IAM mapping, linking GICS sectors to those defined in most IAMs\footnote{To ease the reading, we will refer to sectors used in IAMs as "NGFS sectors"} (which are used in NGFS scenarios and other ones). The objective of this paper is to study the economic and financial shocks induced by climate scenarios in specific sectors and to offer a translation for economic models including numerous variables. While this mapping links material flows (mainly energy and production) from NGFS scenarios to GICS sectors, it does not focus on capturing GHG emissions materiality.
This is precisely the first objective of this paper: to link NGFS and GICS sectors by explicitly incorporating GHG emissions materiality. This approach brings a new dimension, as it captures the physical materiality of GHG emissions and provides specific guidance on handling multiple sectoral correspondences, rather than relying solely on financial indicators or semantic matching. This work represents the first step of this paper.

The second step declines a Sectoral Decarbonization Approach. As mentioned, this top down approach provides sector-specific GHG trajectories consistent with long-term net-zero pathways. Numerous studies have been carried out on the subject, such as the PACTA\footnote{Paris Agreement Capital Transition Assessment} method\footnote{\href{https://rmi-pacta.github.io/r2dii.analysis/articles/target-sda.html}{SDA in the PACTA methodology}} or the Carbon Performance Automobile Manufacturers methodology\footnote{\href{https://researchonline.lse.ac.uk/id/eprint/137832/1/2018-02_Autos_methodology_v1.pdf}{Carbon Performance Automobile Manufacturers Methodology Note}}. The main complexity is to decline this approach when the portfolio and long-term scenarios use different sector classifications. Sectoral mappings can address this issue. Once GICS-NGFS mapping is available, practitioners can include sector-specified GHG emissions materiality into their internal works (such as stress tests and long term projections) with limited amount of data. Here, we will use public data from NGFS climate scenarios and assets spreadsheet from data providers. This paper will provide an illustration of this approach, with a projection of the financed GHG emissions attributed to chosen indices, with a sectoral and geographical breakdown. The results will show how our GICS-NGFS sectoral joint helps to identify the most emissive sectors across time and geographies, to offer better insight on climate impacts and future portfolio management.

This paper is organized as follows: section 2 will focus on the framework used to set the GICS-NGFS mapping. It reviews the existing literature on sectoral mappings, explains the data and methodology, introduces the resulting mapping table, discusses the main limitations of the approach and outlines potential applications and avenues for future research. Section 3 will focus on the Sectoral Decarbonization Approach used to provide GICS sector-specific GHG trajectories from NGFS climate scenarios thanks to the GICS-NGFS mapping. It reviews the existing literature on the SDA approach and explains the data and methodology. Then, the methodology is applied on two illustrative portfolios, represented by indices: MSCI USA and Bloomberg Euro Corporate Bond\footnote{This methodology can also be used for any other indexes, provided that the necessary data explained in section 2 are available. The authors have access to such data for the two used in this paper}. Finally, limitations of the methodology are discussed.

The mapping table tool is publicly available on the Pladifes research program's GitHub\footnote{https://github.com/Pladifes/GICS-NGFS-sectoral-mapping}.

%% file: 2a_NGFS_GICS_jointure.tex
\noindent The NGFS--GICS mapping for GHG emissions proposed in this article builds on two existing frameworks: a GICS--NACE mapping developed by the Banque de France and an IAM--CPRS--NACE mapping introduced by \cite{iam_nace}.

\subsection{Existing literature}

\paragraph{GICS--NACE mapping from the Banque de France} 

For its 2020 climate pilot exercise, the Banque de France developed a correspondence table between GICS and NACE classifications. This initiative aimed to support French banks and insurance companies in assessing climate-related risks through sectoral stress-testing exercises. Since sectoral climate shocks were initially designed based on the sectors of the World Input-Output Database\footnote{The WIOD project is no longer maintained. Archived documentation and data remain available at https://www.rug.nl/ggdc/valuechain/wiod/.}, themselves aligned with the NACE classification, this mapping provided a practical bridge for institutions relying on GICS-based exposures.

\paragraph{IAM--CPRS--NACE mapping}
In 2022, \citeauthor{iam_nace} proposed an IAM--CPRS--NACE mapping framework designed to facilitate the integration of climate scenarios (including NGFS scenarios) into climate-related financial disclosures and stress-testing exercises. The framework aims to connect standard economic activity classifications with the energy and material flow representations used in Integrated Assessment Models (IAMs). As such, it provides a methodological bridge between economic sectors and the transition pathways embedded in climate scenarios.

\paragraph{Extension of the IAM--CPRS--NACE mapping to GICS sectors}
Building on this work, \citeauthor{bat2026} extended the mapping framework to GICS and IEA classification systems in 2026. This extension eases the application of climate scenarios in investment analysis and financial market assessments. The framework preserves the original objective of linking economic activities with energy and material flow variables while allowing users to translate NGFS scenario outputs into GICS-based sectoral exposures.

\paragraph{Research gap} Existing mappings focus on connecting economic or financial classifications with the energy and material flows represented in climate scenarios. Although these frameworks are valuable for climate stress testing and transition risk assessments, they were not explicitly designed to address the allocation of greenhouse gas (GHG) emission trajectories from NGFS scenarios to financial sector classifications. 

\paragraph{Contribution} Our mapping adopts a similar cross-classification approach but focuses specifically on the allocation and evolution of GHG emissions across GICS sectors. By linking NGFS sectoral emission pathways directly to the GICS classification, it enables the assessment and monitoring of financed emissions within investment portfolios under alternative climate scenarios. 

\begin{figure}[H]
    \centering
    \includegraphics[width=0.8\textwidth]{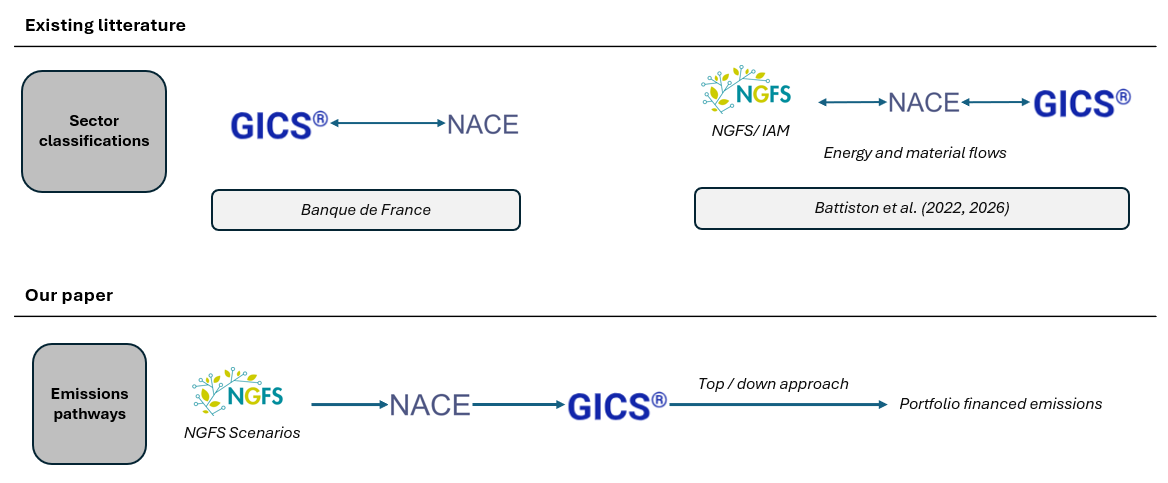}
    \caption{\footnotesize Existing mapping frameworks and proposed NGFS--GICS emissions mapping. The upper panel summarizes the existing literature linking climate scenarios and sector classifications. The lower panel presents the contribution of this paper, which maps NGFS emission pathways to GICS sectors and enables portfolio-level financed-emissions analysis through a top-down approach.}
    \label{fig:literature_mapping}
\end{figure}

\subsection{General methodology}

\paragraph{Prerequisites: mapping structures}

Although the primary objective is to establish correspondences at an aggregated sectoral level, examining the classifications' sub-sectors is needed to improve the accuracy and transparency of the mapping process. For this reason, the different levels of granularity of the GICS, NACE, and NGFS classifications are explicitly distinguished throughout the analysis.

The GICS classification comprises four hierarchical levels: sectors, industry groups, industries, and sub-industries. Similarly, the NACE classification is organized into four levels: sections, divisions, groups, and classes. Although NGFS scenarios do not formally define sectoral granularity levels, several variables are reported at a more detailed sub-sector level. Consequently, we distinguish between global sectors and sub-sectors within the NGFS framework.

For consistency across classifications, we introduce a common notation based on four levels of granularity, ranked from the most aggregated to the most detailed: \textit{Classification}-1, \textit{Classification}-2, \textit{Classification}-3, and \textit{Classification}-4. For example, the industry group level of the GICS classification is denoted as \textit{GICS-2}.

\paragraph{Methodology}

\noindent The objective of this mapping is to link NGFS and GICS sectors based on the materiality of their Scope 1 and Scope 2 emissions, rather than on a purely semantic correspondence. More specifically, the Scope 1 and Scope 2 emissions associated with each NGFS-2 category are allocated to one or several GICS-2 categories according to the underlying economic activities and their GHG emission profiles.

The mapping was developed through the following steps:
\begin{enumerate}
    \item Each NGFS-1 sector is made up of several NGFS-2 categories. Each NGFS-2 category is linked to one or several NACE-2 categories. Because NGFS sub-sectors are generally broad, they often encompass multiple NACE divisions.
    \item Each identified NACE-2 category is subsequently assigned to the corresponding GICS-2 categories, which are then aggregated to the corresponding GICS-1 sector.
\end{enumerate}

The GICS--NACE mapping (\cite{nace_gics}) and the IAM--CPRS--NACE mapping (\cite{iam_nace}) were used as starting points for this process, provided that the underlying correspondences were consistent with the objective of matching sectors according to the materiality of their Scope 1 and Scope 2 emissions. Such correspondences were considered sufficiently robust for approximately 77\% of total GHG emissions (from NGFS data). The remaining NGFS-2 categories required a case-by-case assessment to ensure that the resulting mapping reflected the materiality of Scope 1 and Scope 2 emissions rather than purely semantic or activity-based similarities.

\paragraph{Identification and treatment of specific cases}

The main adjustments concern sectors for which economic-activity classifications and GHG emission profiles diverged. Details are provided in Appendix A.

%% file: 2b_Results.tex
\subsection{Results}

Our work results in a mapping table between NGFS and GICS sectors, with NACE codes as main conversion key. Each column represents a sectoral level: \\

\begin{tabular}{|l|p{12cm}|} 
    \hline 
    \textbf{Code} & \textbf{Description} \\ 
    \hline NGFS-1 & The most global level of NGFS sectors \\ 
    \hline NGFS-2 & The NGFS sub-sectors \\ 
    \hline NACE code & The NACE division (the second most global NACE classification level) \\ 
    \hline NACE description & A quick description of the NACE division \\ 
    \hline GICS-1 & The most global level of GICS sectors \\ 
    \hline GICS-2 & GICS industry group (the second most global GICS classification level) \\ 
    \hline 
\end{tabular} 

\begin{figure}[H]
  \centering
  \includegraphics[scale = 0.45]{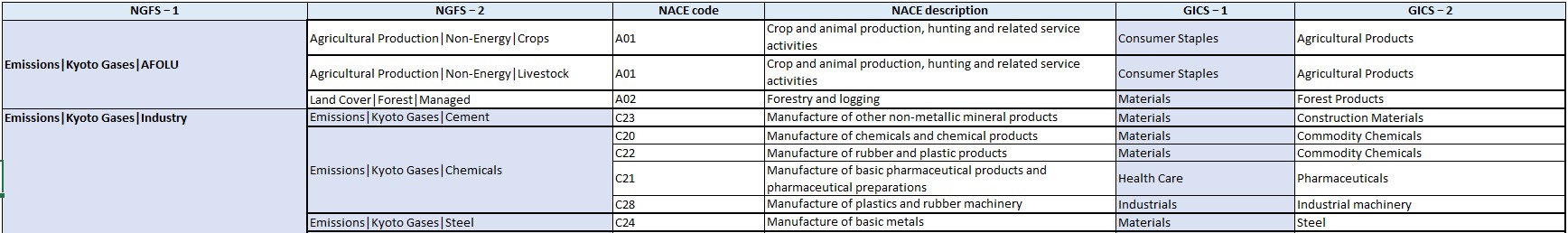}
  \caption{Extract from the GICS-NGFS mapping}
  \label{fig:center}
\end{figure}   

This mapping is saved in an Excel file (\texttt{GICS-NGFS-GHG-mapping.xlsx}), available on the Pladifes.

%% file: 2c_Discussion_conclusion.tex
\subsection{Summary and limitations}

\noindent This proposed GICS-NACE sectoral mapping provides a tool to assist financial professionals in assessing their climate impact. It meets the need to fine-tune climate investment strategies and climate risk exposure estimation by joining classification systems designed solely for the purpose of classifying activities or quantifying physical flows. The present mapping offers a link between the GICS and the NGFS emissive sectors, to translate NGFS climate scenarios in portfolio specific projections at different time horizons, for diverse narratives. The tool can be extended to other sectors and classifications (such as the IEA) by applying the same logic of mapping physical inputs to GHG emissions. Its current state is usable by investors and portfolio managers, but it can be developed and fine-tuned for increased precision.

Some limitations on the proposed mapping can be pointed out. This tool focuses only on the materiality of the GHG and not other material flows such as energy, defining a clear perimeter for its use. Also, the "Other Industry" NGFS sub-sector could potentially be linked to more GICS sectors, as numerous GICS industry groups may fit the joint. The joint relies mainly on sector descriptions, with GHG emission flows that may not correspond exactly across classifications. Naturally, the application displayed the dependence to NGFS GHG data, which are derived from specific scenario and model hypotheses. This limit is inherent to any climate scenario provider and can hardly be bypassed.

%% file: 3a_Materials_and_methods.tex
\subsection{Objective and existing literature} 
The methodology aims to estimate portfolio-specific Scope 1 and 2 emission trajectories at GICS-1 level sector granularity. This approach uses NGFS long term climate scenarios, portfolio composition data, and the GICS-NGFS mapping described previously.

Some works have been conducted on sectoral GHG projections using top down approaches: \citeauthor{vashold} (\citeyear{vashold}) developed a model combining sectoral Kaya identities, auto-regressive vector models, and Bayesian estimations to define GHG projections of five sectors based on SSP trajectories. \citeauthor{janssens} (\citeyear{janssens}) proposed estimations between GHG emissions and explanatory factors to produce projections based on climate scenarios.

\subsection{Necessary data used}
\paragraph{NGFS long term scenarios} The methodology relies on the NGFS variable \texttt{GCAM|\allowbreak\ Emissions|Kyoto Gases}, which provides the trajectories of Kyoto Protocol greenhouse gas emissions generated using the GCAM\footnote{Global Change Analysis Model} integrated assessment model. These trajectories are available for each NGFS region and sector and are expressed in carbon dioxide equivalent (\(CO_2e\)). The proposed framework can be applied to any NGFS climate pathway generated by GCAM.

Among the three integrated assessment models provided within the NGFS framework, namely GCAM, REMIND-MAgPIE and MESSAGEix-GLOBIOM, GCAM was selected because it offers the broadest coverage of sector-level greenhouse gas emissions expressed in \(CO_2e\). Although REMIND-MAgPIE and MESSAGEix-GLOBIOM provide a finer sectoral disaggregation, their coverage of Kyoto-gas emissions is more limited. Since the objective of the analysis is to estimate trajectories for all greenhouse gases covered by the Kyoto Protocol rather than for carbon dioxide emissions alone, the availability of \(CO_2e\)-based variables constitutes a key selection criterion.

\paragraph{Portfolio composition data} The methodology requires the following inputs :
\begin{itemize} 
    \item the distribution of portfolio market capitalization across NGFS regions (\%). Since portfolio exposures are generally reported at the country level, the NGFS regional mapping is used to aggregate countries into NGFS regions; 
    \item the distribution of Scope 1 and Scope 2 greenhouse gas emissions (\%) across GICS-1 sectors. 
\end{itemize} 

\subsection{Portfolio emissions projection framework} 

The following figure summarizes the different steps of the proposed methodology, from NGFS climate scenarios to portfolio-level emissions trajectories.

\input{method_schema}

Let $E^{\mathrm{Portfolio}}(N)$ denote the portfolio Scope 1 and 2 emissions at a reference year $N$. The projected portfolio emissions at year $t$ are defined as 
\begin{equation} 
    E^{\mathrm{Portfolio}}(t) = E^{\mathrm{Portfolio}}(N)\times DF^{\mathrm{Portfolio}}(t)
\end{equation} 
where $DF^{\mathrm{Portfolio}}(t)$ represents the portfolio decarbonization factor relative to the reference year $N$. 

The portfolio decarbonization factor is computed as a weighted average of GICS-1 sector decarbonization factors: 
\begin{equation} 
    DF^{\mathrm{Portfolio}}(t) = \sum_{j} W_j^{\mathrm{GICS}} \, DF_j^{\mathrm{GICS}}(t)
\end{equation}
where $W_j^{\mathrm{GICS}}$ denotes the the share of portfolio Scope 1+2 emissions attributable to GICS-1 sector $j$, and $DF_j^{\mathrm{GICS}}(t)$ is the corresponding sectoral decarbonization factor. 

\subsubsection{GICS-1 sector weights \texorpdfstring{$W^{\mathrm{GICS}}_j$}{W GICS j}}

The weight associated with GICS-1 sector $j$ is defined as the share of portfolio Scope 1+2 emissions attributable to companies operating in this sector in the reference year $N$ : 
\begin{equation}
    W^{\mathrm{GICS}}_j = \frac{\sum_{k \in j} E_{k}(N)}{\sum_k E_k(N)} 
\end{equation}
where $E_k(N)$ denotes the scope 1+2 emissions of company $k$ in the reference year $N$.

\paragraph{Why use emissions-based weights?} 
\label{par:emissions_weights}
Sector weights are based on Scope 1+2 emissions rather than market capitalization. This approach ensures that the allocation reflects the portfolio's actual GHG emission profile. A market-capitalization-based approach would require the use of sector-level economic intensities and would be sensitive to changes in asset prices and market conditions. As a result, sector weights could vary over time even when underlying emissions remain unchanged. By contrast, emissions-based weights rely on observed company emissions and follow a bottom-up aggregation approach are used. They therefore provide a more direct measure of each sector's contribution to the portfolio carbon footprint and offer a more appropriate basis for applying sectoral decarbonization pathways.

\paragraph{Why assume constant emissions weights?}
The weights are assumed to remain constant throughout the projection horizon. While portfolio compositions evolve over time, reliable forward-looking information on future sector allocations is generally unavailable. The sector distribution observed in the reference year $N$ is therefore used as a proxy for future portfolio composition. This assumption allows the analysis to isolate the impact of sectoral decarbonization pathways on portfolio emissions, independently of future portfolio re-balancing effects.

\subsubsection{Sectoral Decarbonization Factors \texorpdfstring{$DF^{\mathrm{GICS}}_j(t)$}{DF[GICS]j(t)}} 
A given GICS-1 sector may be associated with several NGFS sectors and geographical regions. The sectoral decarbonization factors used in the portfolio projection framework are therefore obtained through a multi-stage aggregation process combining NGFS climate scenarios, the GICS-NGFS mapping and geographical adjustment factors.

\paragraph{Step 1. NGFS decarbonization factors}

For each NGFS region and NGFS-2 sector combination $(k,l)$, a decarbonization trajectory is derived from the NGFS variable \texttt{GCAM|Emissions|Kyoto Gases}. The decarbonization factor at year $t$ relative to the reference year $N$ is defined as
\begin{equation}
    DF^{\mathrm{NGFS}}_{k,l} (t) = \frac{NGFS_{k,l}(t)}{NGFS_{k,l}(N)}
\end{equation}
where $NGFS_{k,l}(t)$ denotes the GCAM Kyoto Gases emissions associated with NGFS region $k$ and NGFS-2 sector $l$ for the year $t$\footnote{GCAM scenarios are reported at five-year intervals. Consequently, the selected reference year $N$ may not be directly available in the dataset. In such cases, emissions for year $N$ are estimated using linear interpolation between the two surrounding NGFS observations.}.

\paragraph{Step 2. Geographical adjustment} 
Because the same NGFS sector may exhibit very different emissions trajectories across regions, regional trajectories are weighted according to both their relative emissions level and the geographical composition of the portfolio. 
Let $W_k^{\mathrm{Region}}$ denote the share of portfolio market capitalization allocated to the NGFS region $k$. The geographical adjustment factor is defined as 
\begin{equation} 
    GF_{k,l} = W_k^{\mathrm{Region}} \, \frac{NGFS_{k,l}(N)} {\sum_k NGFS_{k,l}(N)}. 
\end{equation} 
This factor reflects both the geographical exposure of the portfolio and the relative importance of the region $k$ within the NGFS sector $l$. This factor is assumed to remain constant throughout the projection horizon.

\paragraph{Step 3. Scope 1+2 adjustment factor} NGFS GHG emissions data include emissions associated with electricity generation. Since Scope 2 emissions already account for indirect emissions arising from electricity consumption, a direct use of NGFS data could lead to double counting. To mitigate this issue, the data is adjusted using the following factor: 
\begin{equation} 
    F_{adj,k} = \frac{E^{tot}_k} {E^{tot}_k + E^{elec}_k}, 
\end{equation} 
where $E^{tot}_k$ denotes the total emissions in the NGFS region $k$, and $E^{elec}_k$ corresponds to the emissions from electricity generation. The adjustment factor is computed independently for each NGFS region and is assumed to remain constant throughout the projection horizon.

\paragraph{Step 4. Allocation of NGFS emissions to GICS sectors} 
NGFS emissions trajectories are reported at the NGFS-2 sector level, whereas portfolio exposures are measured using the GICS-1 classification. Consequently, NGFS trajectories must be translated into GICS-based trajectories using the mapping presented in Section~\ref{sec:GICS-NGFS mapping}. 
For a given NGFS-2 sector $l$, the emissions are first allocated to the associated GICS-1 sectors according to their relative contribution to their emissions in Scope 1 and 2. The corresponding allocation weight is defined as 
\begin{equation} 
    w_{l,j}^{\mathrm{GICS}\rightarrow\mathrm{NGFS}} = \frac{E_j^{\mathrm{GICS}}} {\sum_{m \in l} E_m^{\mathrm{GICS}}} 
\end{equation} 
where $E_j^{\mathrm{GICS}}$ denotes the emissions in Scope 1 and 2 associated with the GICS-1 sector $j$, and the denominator aggregates emissions in all GICS-1 sectors mapped to the NGFS-2 sector $l$. These weights are subsequently applied to NGFS emissions in order to derive indicative emissions for each NGFS region-sector pair $(k,l)$ and GICS-1 sector $j$: 
\begin{equation} 
    E_{k,l,j}^{\mathrm{ind}} = w_{l,j}^{\mathrm{GICS}\rightarrow\mathrm{NGFS}} \times F_{adj,k} \times NGFS_{k,l}(N), 
\end{equation} 
where $NGFS_{k,l}(N)$ denotes NGFS emissions for region $k$ and sector $l$ in the reference year $N$, and $F_{adj,k}$ is the adjustment factor of Scope 1 and 2 for region $k$. The resulting indicative emissions are then used to estimate the contribution of each NGFS region-sector pair $(k,l)$ to the decarbonization trajectory of GICS-1 sector $j$. This contribution is defined as 
\begin{equation} 
    w_{k,l,j}^{\mathrm{NGFS}\rightarrow\mathrm{GICS}} = \frac{E_{k,l,j}^{\mathrm{ind}}} {\sum_{k,l} E_{k,l,j}^{\mathrm{ind}}}. 
\end{equation} 

\paragraph{Step 5. Aggregation into GICS sector decarbonization rates} 

The decarbonization rate associated with the GICS sector $j$ is finally obtained by aggregating the NGFS decarbonization factors across all associated combinations of the NGFS region-sector: 
\begin{equation} 
    DF_j^{\mathrm{GICS}}(t) = \sum_{k,l} w_{k,l,j}^{NGFS \rightarrow GICS} \, GF_{k,l} \, DF_{k,l}^{\mathrm{NGFS}}(t). 
\end{equation} 
The resulting trajectory represents the decarbonization pathway assigned to the GICS-1 sector $j$, which accounts for both mapping of the NGFS-GICS sector and the geographical distribution of the portfolio across the NGFS regions.

%% file: method_schema.tex
\begin{figure}[htbp] \centering \resizebox{\textwidth}{!}{%
\begin{tikzpicture}[ font=\small, box/.style={ rectangle, rounded corners, draw, thick, minimum width=4cm, minimum height=1cm, align=center }, header/.style={ rectangle, draw, fill=gray!20, thick, minimum width=5cm, minimum height=0.8cm, font=\bfseries }, arrow/.style={->, thick} ] 
\node[header] (inputs) at (0,4) {INPUTS}; \node[header] (method) at (8,4) {METHOD}; \node[header] (outputs) at (16,4) {OUTPUTS}; 
\node[box] (ngfs) at (0,2) {NGFS Climate Scenarios\\ \textit{(GCAM | Kyoto Gases)}}; \node[box] (mapping) at (0,0) {GICS--NGFS\\ Mapping}; \node[box] (regional) at (0,-2) {Portfolio Regional Allocation}; \node[box] (weights) at (0,-4) {Portfolio Scope1+2 emissions}; 
\node[box] (gics) at (8,0) {GICS-1 Sector\\ Emissions Pathways \\ \small{\textit{(Scope 1+2 and regionally adjusted)}}};  \node[box] (weighting) at (8,-4) {Portfolio Emissions Weighting}; 
\node[box] (portfolio) at (16,-2) {Projected Portfolio\\ Emissions Pathway}; 
\draw[arrow] (ngfs.east) -- (gics.west); \draw[arrow] (mapping.east) -- (gics.west); \draw[arrow] (regional.east) -- (gics.west); \draw[arrow] (weights.east) -- (weighting.west); 
\draw[arrow] (gics.east) -- (portfolio.west); \draw[arrow] (weighting.east) -- (portfolio.west);; 
\end{tikzpicture} } 
\caption{Overview of the portfolio emissions projection framework.} 
\label{fig:framework_schema} 
\end{figure}
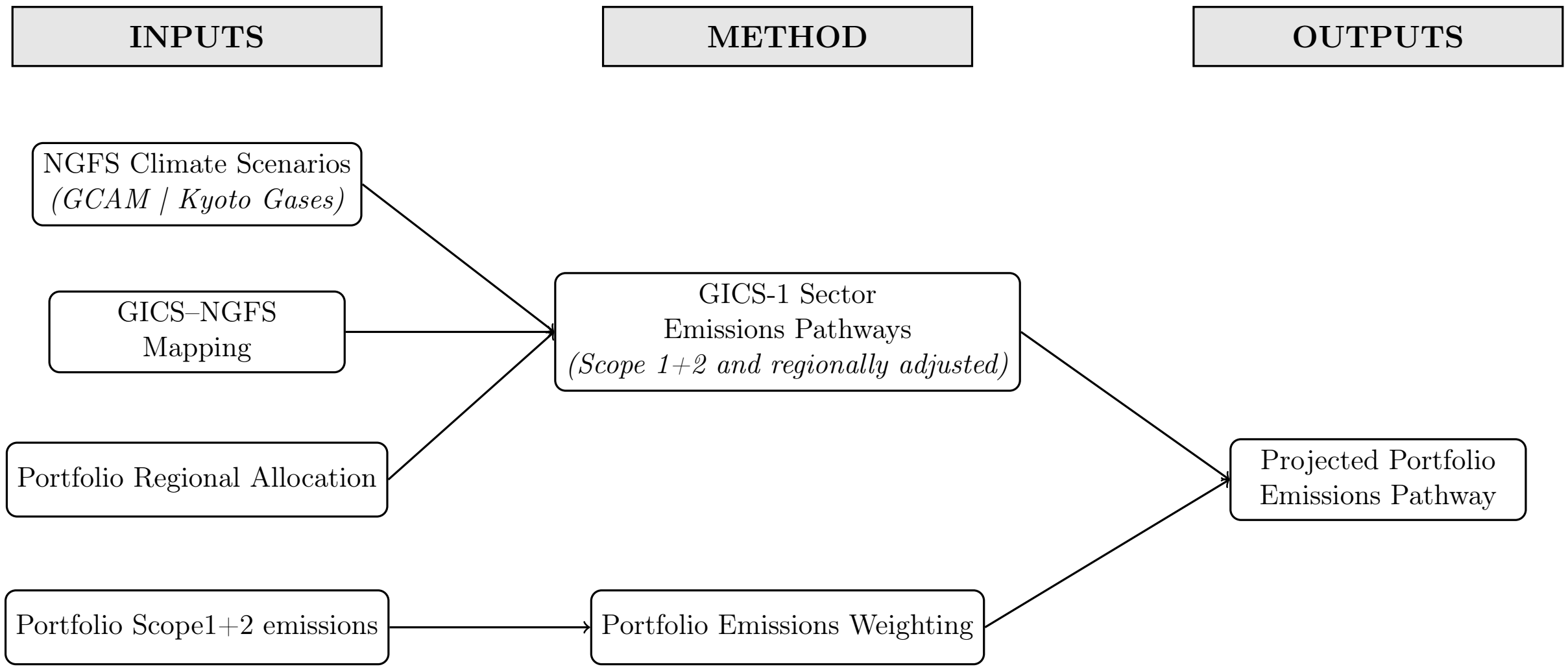

%% file: 3b_Application.tex
\subsection{Applications on two ETFs}

This subsection illustrates the proposed methodology through a practical case study involving two ETFs tracking the MSCI USA Index and the Bloomberg Euro Corporate Bond Index.

\subsubsection{Data used} 

\paragraph{NGFS GCAM scenarios} We used phase V of the NGFS GCAM scenarios released on November 5, 2024\footnote{\url{https://data.ene.iiasa.ac.at/ngfs-phase-5/}}. We focus on two contrasted scenarios : Net Zero 2050 (most orderly transition pathway) and Current Policies (business-as-usual pathway, least orderly).

\paragraph{ETF Characteristics} \hfill\mbox{}
\begin{table}[H] 
\centering 
    \begin{tabular}{|p{3cm}|m{6cm}|m{7.5cm}|} 
        \hline & {\centering \textbf{MSCI USA}} & {\centering \textbf{Bloomberg Euro Corporate Bond}} \\ 
        \hline Asset type & Stocks & Bond Investment Grade \\ 
        \hline ETF full name and link & \href{https://www.blackrock.com/fr/intermediaries/products/253740/ishares-msci-usa-b-ucits-etf}{iShares MSCI USA UCITS ETF} & \href{https://www.ishares.com/uk/individual/en/products/251726/ishares-euro-corporate-bond-ucits-etf}{iShares Core EUR Corp Bond UCITS ETF} \\ 
        \hline Extraction date & January 2025 & January 2025 \\ 
        \hline Market capitalization decomposition according to NGFS region & {\centering \includegraphics[width=6cm]{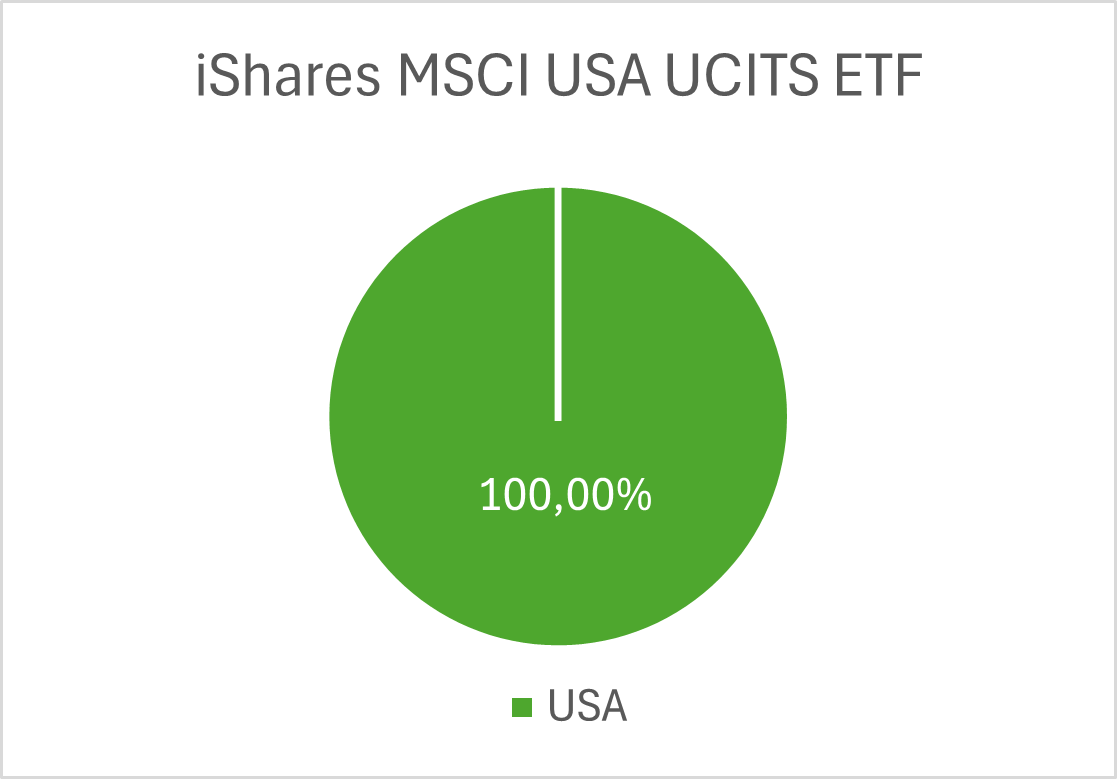}} & {\centering \includegraphics[width=7.5cm]{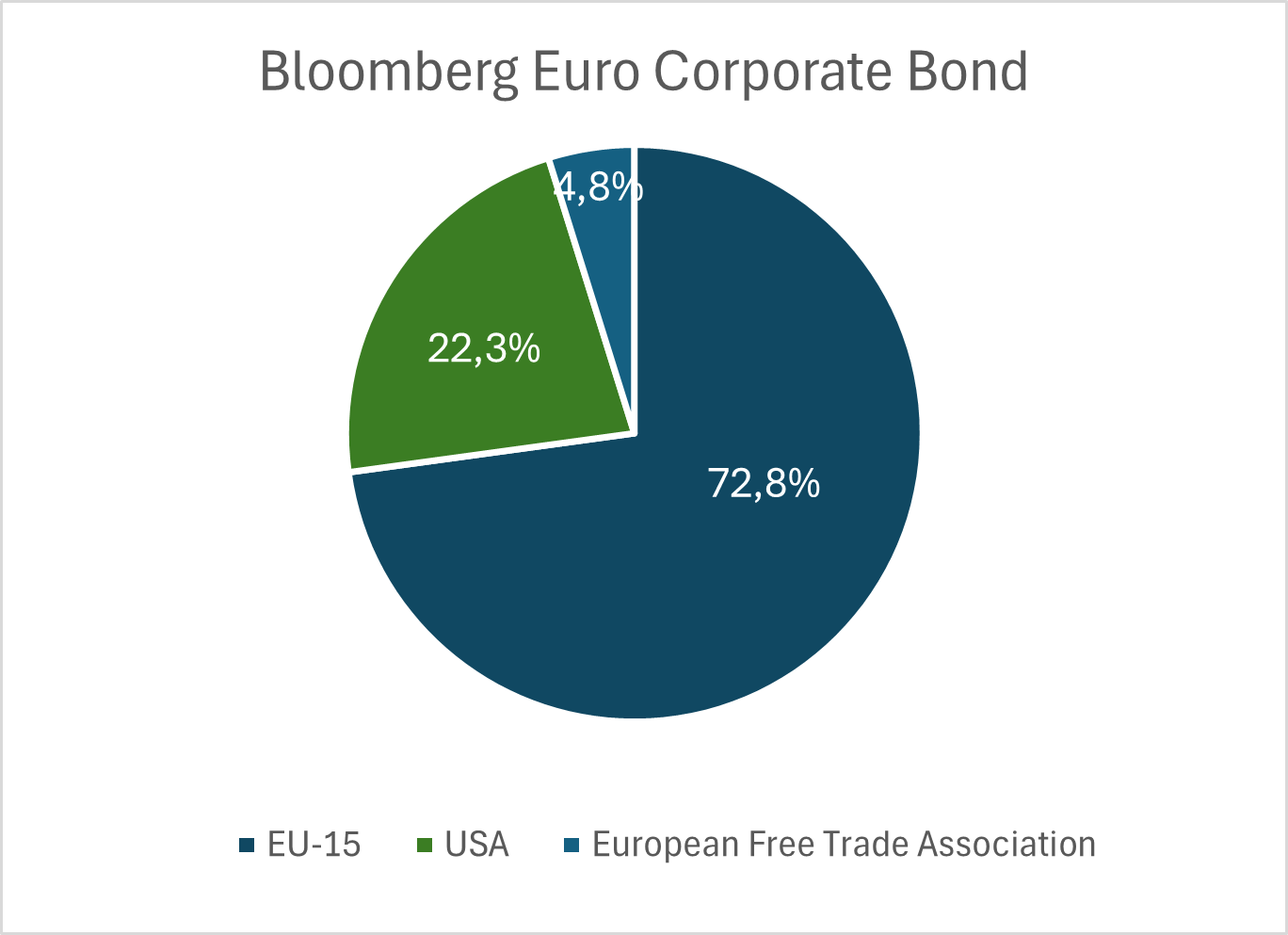}} \\
        \hline
    \end{tabular} 
\caption{Data source of the ETFs} 
\label{tab:etf_characteristics} 
\end{table}

The two portfolios invest in different asset classes (equities and corporate bonds) and exhibit distinct geographical exposures. For MSCI USA, the NGFS decarbonization factor will be calculated on only one NGFS region (USA). For Bloomberg Euro Corporate Bond, three NGFS regions : USA, EU-15\footnote{Fifteen pre-2004 European Union member states : Austria, Belgium, Denmark, Finland, France, Germany, Greece, Ireland, Italy, Luxembourg, the Netherlands, Portugal, Spain, Sweden, the United Kingdom} and Europe Free Trade Association (EU FTA\footnote{Iceland, Norway and Switzerland.}).

\paragraph{GICS-1 sector Scope 1+2 emissions decompositions}
This particular data is not publicly available. As mentioned in \cref{par:emissions_weights}, it is necessary to retrieve Scope 1+2 emissions data for each company listed within those ETFs. In this case, the COGEM database was used: it offers estimated GHG data for more than 47,000 listed companies, using a transparent methodology\footnote{The COGEM database is part of the PLADIFES research program of Institut Louis Bachelier. More information here: https://pladifes.institutlouisbachelier.org/ghg-emissions/}.

As a result, the following table gathers the Scope 1+2 emissions according to GICS-1 sectors : 

\begin{table}[H] 
\centering 
    \begin{tabular}{|p{6cm}|m{3cm}|m{5cm}|} 
        \hline {\centering \textbf{GICS-1 sectors}} & {\centering \textbf{MSCI USA}} & {\centering \textbf{Bloomberg Euro Corporate Bond}} \\ 
        \hline Communication & 1.21\% & 1.35\% \\ 
        \hline Consumer Discretionary & 2.65\% & 3.91\% \\
        \hline Consumer Staples & 5.00\% & 5.83\% \\
        \hline Energy & 23.94\% & 19.94\% \\
        \hline Financials & 0.23\% & 0.33\% \\
        \hline Health Care & 1.61\% & 0\% \\
        \hline Industrials & 10.31\% & 33.36\% \\
        \hline Information Technology & 2.01\% & 1.16\% \\
        \hline Materials & 20.29\% & 0\% \\
        \hline Real Estate & 0.45\% & 0\% \\
        \hline Utilities & 32.30\% & 34.13\% \\
        \Xhline{1.5pt} TOTAL (MtCO2e) & 1336.48 & 990.41 \\   
        \hline
    \end{tabular} 
\caption{Breakdown of scope 1+2 emissions by GICS-1 sector} 
\label{tab:etf_emissions} 
\end{table}

\subsubsection{Projected Emissions Pathways} 
    The following figure displays the projected emissions pathways of each ETF, considering 2025 as the base year:

    \begin{figure}[htbp] 
        \centering 
        \begin{minipage}{0.48\textwidth} 
            \centering 
            \includegraphics[width=\linewidth]{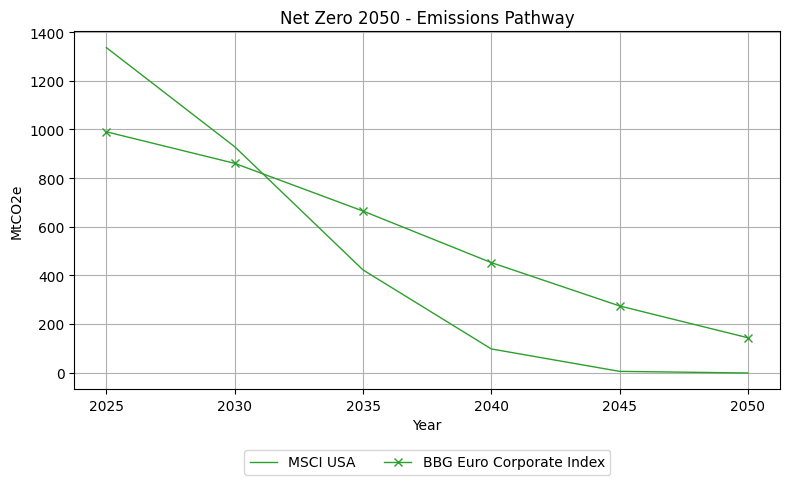} 
        \end{minipage} 
        \hfill 
        \begin{minipage}{0.48\textwidth} 
            \centering 
            \includegraphics[width=\linewidth]{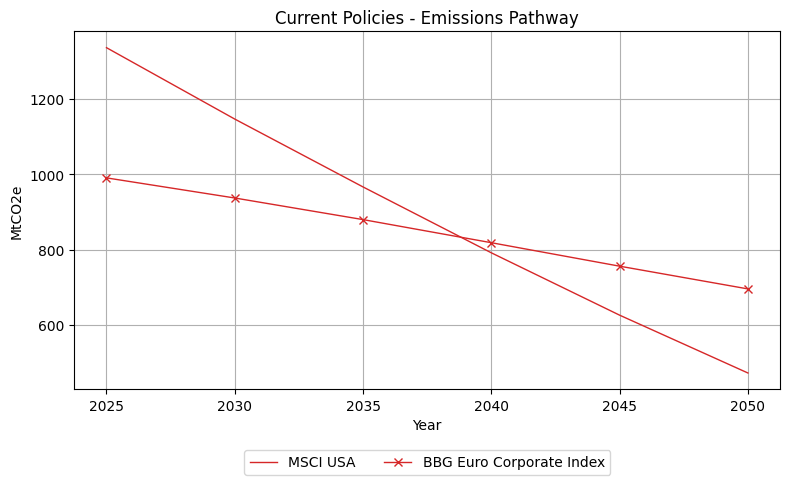}
        \end{minipage} 
    \caption{ETF emissions pathways under Net Zero (left) and Current Policies (right) scenarios}
    \label{fig:NZ_CP} 
    \end{figure}

Despite its higher initial emissions, MSCI USA decarbonizes faster than Bloomberg Euro Corporate Bond for all scenarios. The following figures show their corresponding normalized emissions pathway.

    \begin{figure}[htbp] 
        \centering 
        \begin{minipage}{0.48\textwidth} 
            \centering 
            \includegraphics[width=\linewidth]{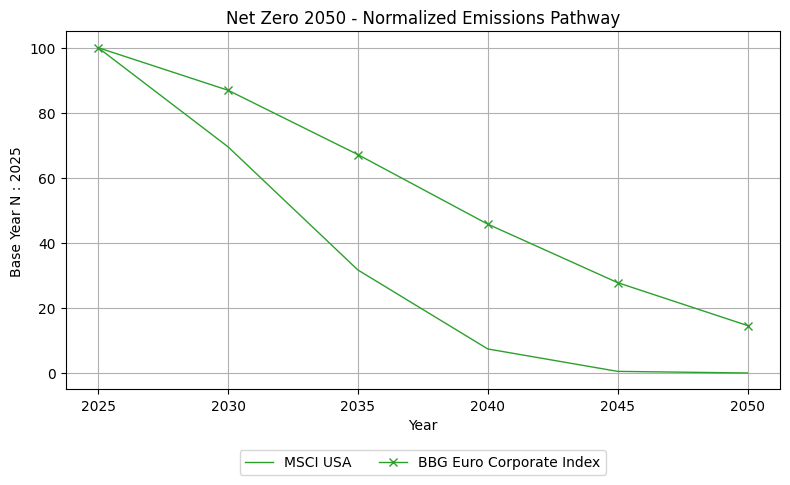} 
        \end{minipage} 
        \hfill 
        \begin{minipage}{0.48\textwidth} 
            \centering 
            \includegraphics[width=\linewidth]{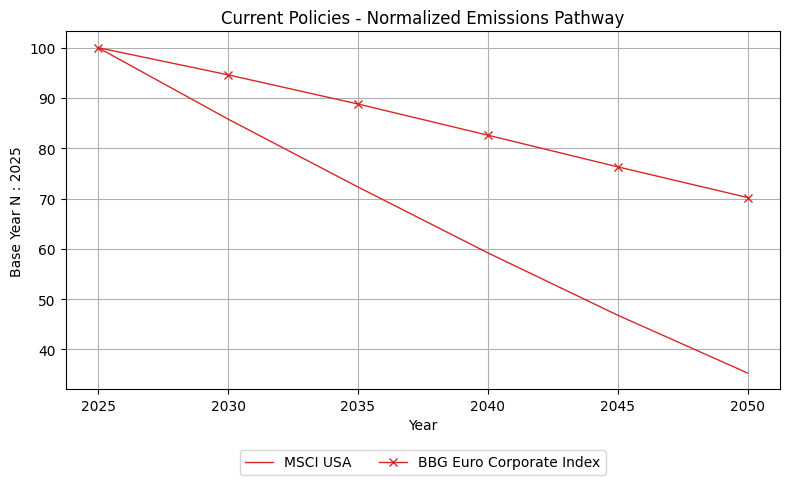}
        \end{minipage} 
    \caption{ETF normalized emissions pathways under Net Zero (left) and Current Policies (right) scenarios}
    \label{fig:NZ_CP2} 
    \end{figure}  

\subsubsection{Identifying the Key Drivers of Divergence} 
The divergence between the two pathways is driven by differences in sectoral and geographical decomposition. The following subsection led an analysis on the Current Policies scenario to identify the main factors of this difference\footnote{The analysis would be similar for Net Zero scenario}. 

To identify the drivers of divergence between the normalized emissions pathways of MSCI USA and the Bloomberg Euro Corporate Bond, we first examine the sectoral composition of portfolio emissions and then decompose the resulting decarbonization factors into their underlying NGFS sector and regional components. More than 85\% of portfolio emissions are concentrated within four GICS-1 sectors: \textit{Energy}, \textit{Industrials}, \textit{Materials}, and \textit{Utilities}. 
\begin{table}[H] 
\centering 
\begin{tabularx}{\textwidth}{|X|c|c|} 
\hline 
    \textbf{GICS Level 1 Sector} & \textbf{MSCI USA} & \textbf{Bloomberg Euro Corporate Bond} \\ 
    \hline Energy & 23.94\% & 19.94\% \\ 
    Industrials & 10.31\% & 33.36\% \\ 
    Materials & 20.29\% & 0\% \\ 
    Utilities & 32.30\% & 34.13\% \\ 
    \hline 
\end{tabularx} 
\caption{Emission weights of the main GICS Level 1 sectors} 
\label{tab:gics_weights} 
\end{table} 

\begin{figure}[htbp] 
\centering 
\begin{minipage}{0.48\textwidth}            
    \includegraphics[width=\linewidth]{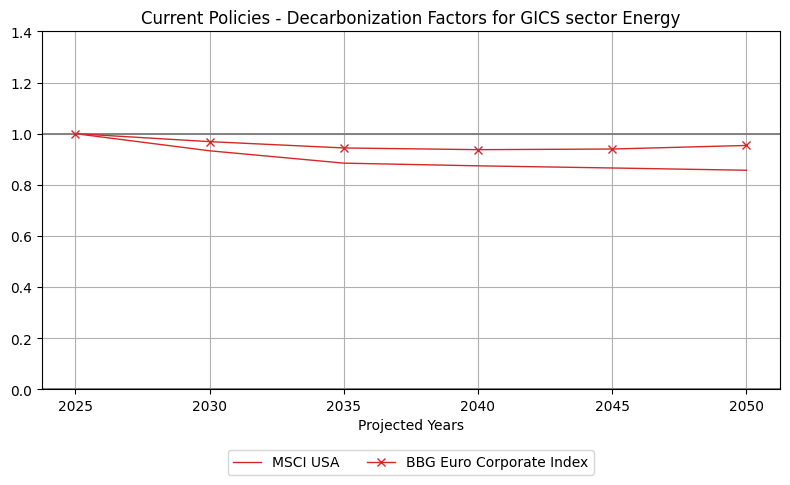}
\end{minipage} 
\hfill 
\begin{minipage}{0.48\textwidth} 
    \includegraphics[width=\linewidth]{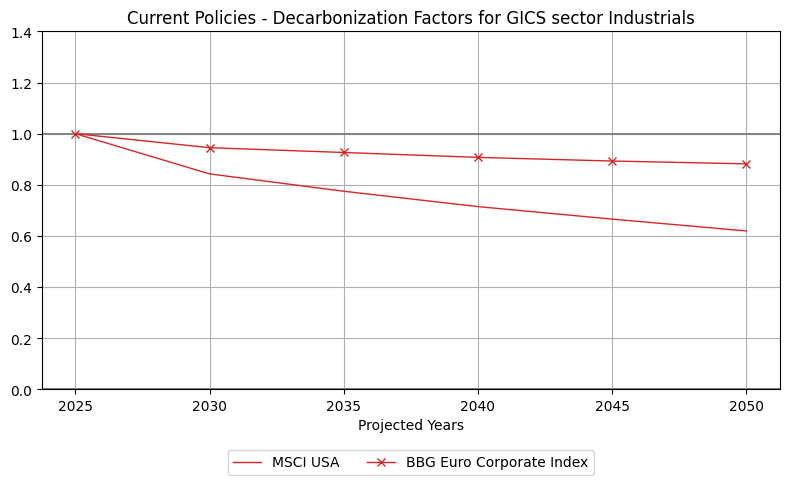}
\end{minipage} 
\vspace{0.3cm} 
\begin{minipage}{0.48\textwidth} 
    \includegraphics[width=\linewidth]{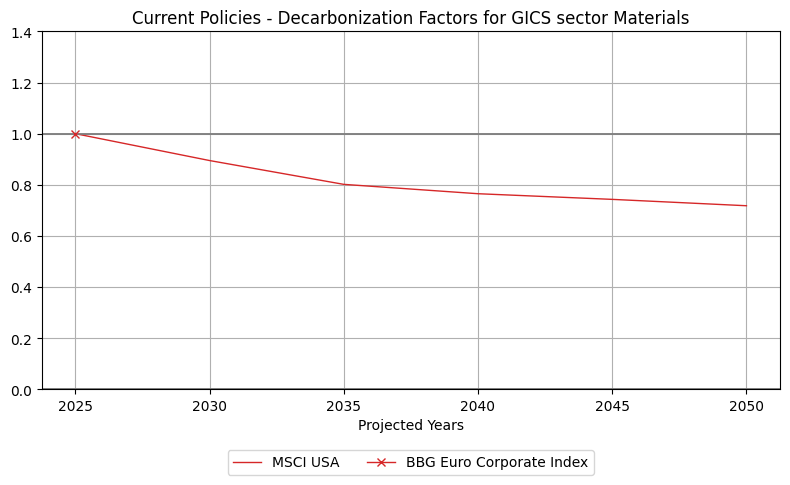}
\end{minipage} 
\hfill 
\begin{minipage}{0.48\textwidth} 
    \includegraphics[width=\linewidth]{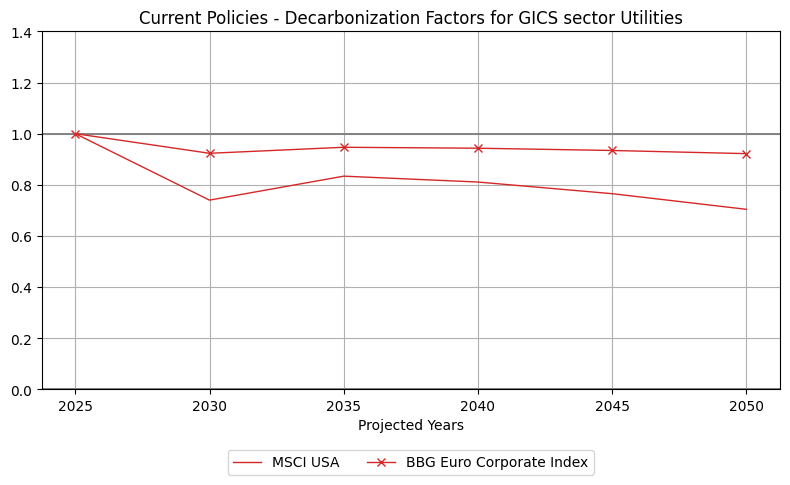} 
\end{minipage}
    \caption{Decarbonization factors of the most material GICS-1 sectors}
    \label{fig:gics_contributions} 
\end{figure}

Figure~\ref{fig:gics_contributions} compares the corresponding GICS-1 sector decarbonization factors.

Two sectors can be excluded as major sources of divergence. First, the \textit{Energy} sector exhibits similar emission weights and decarbonization factors across both ETFs, resulting in nearly identical contributions to the normalized pathways. Second, although the \textit{Industrials} sector exhibits a stronger decarbonization factor in MSCI USA, this effect is largely diminished by its substantially lower emission weight relative to the Bloomberg Euro Corporate Bond. 

By contrast, \textit{Materials} and \textit{Utilities} sectors emerge as the principal drivers of divergence. \textit{Materials} is only represented in MSCI USA. Combined with a pronounced decarbonization factor, it contributes significantly to the accelerated decrease of the MSCI USA emissions pathway. \textit{Utilities} exhibits comparable emission weights in both ETFs but noticeably different decarbonization factors, causing a visible divergence in sector contributions. To understand these differences, the two sectors are decomposed into their underlying NGFS-2 sectors. 

\begin{table}[H] 
\centering 
\small
\begin{tabularx}{\textwidth}{|X|c|c|c|} 
\hline \textbf{GICS-1 Sectors} & \textbf{NGFS-2 Sectors} & \textbf{MSCI USA} & \textbf{Bloomberg Euro Corporate Bond} \\ 
\hline \multirow{5}{*}{\textbf{Materials}} & Other Industry & 32.57\% & \\ 
& Chemicals & 23.25\% & \\ 
& Other Energy Supply & 19.14\% & \\ 
& Steel & 12.38\% & \\ 
& Cement & 12.66\% & \\ 
\hline \multirow{2}{*}{\textbf{Utilities}} & Electricity & 90.36\% & 84.90\% \\ 
& Other Energy Supply & 9.64\% & 15.10\% \\ 
\hline 
\end{tabularx} 
\caption{Breakdown of selected GICS sectors into NGFS-2 sectors} 
\label{tab:ngfs_breakdown} 
\end{table}

For the \textit{Materials} sector, five NGFS-2 sectors contribute to the overall decarbonization factor. \begin{figure}[H] 
\centering 
\includegraphics[width=\linewidth]{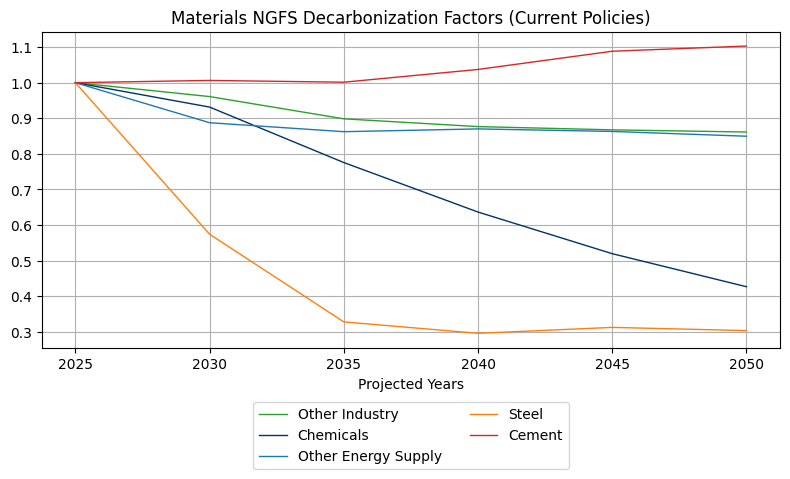} 
\caption{NGFS sector-level contributions to the \textit{Materials} decarbonization factor} 
\label{fig:materials_ngfs} 
\end{figure} 

The results show that the decarbonization factor is primarily driven by the \textit{Other Industry}, \textit{Chemicals} and \textit{Steel} sectors. Because the \textit{Materials} sector is absent from the Bloomberg Euro Corporate Bond, these trajectories contribute exclusively to the MSCI USA emissions pathway and constitute a first source of divergence.

Unlike \textit{Materials} sector, which is characterized by a sector-composition effect, the divergence observed for \textit{Utilities} originates from geographical differences within the same sector. The decomposition reveals that the \textit{Electricity} NGFS-2 sector alone explains more than 85\% of the \textit{Utilities} decarbonization factor for each ETF.

\begin{figure}[H] 
\centering 
\includegraphics[width=0.9\textwidth]{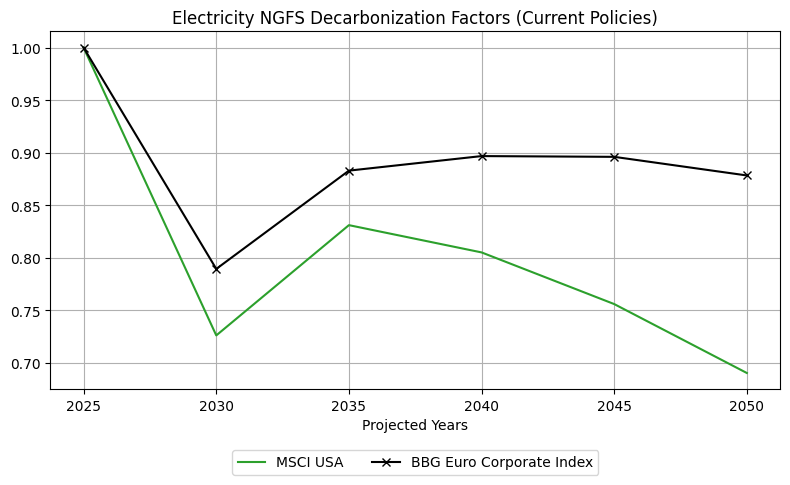} 
\caption{\textit{Electricity} NGFS decarbonization factor for both ETFs} 
\label{fig:electricity_total} 
\end{figure} 

\begin{figure}[H] 
\centering 
\begin{minipage}{0.48\textwidth} 
\centering 
\includegraphics[width=\linewidth]{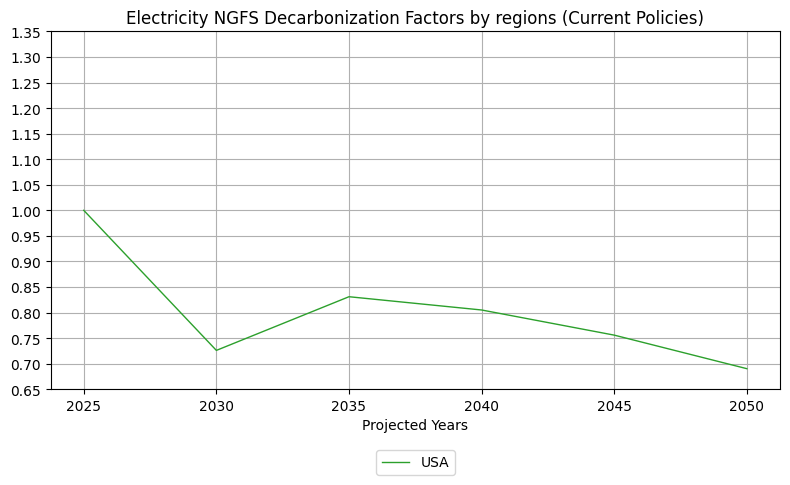} \end{minipage} 
\hfill 
\begin{minipage}{0.48\textwidth} 
\centering 
\includegraphics[width=\linewidth]{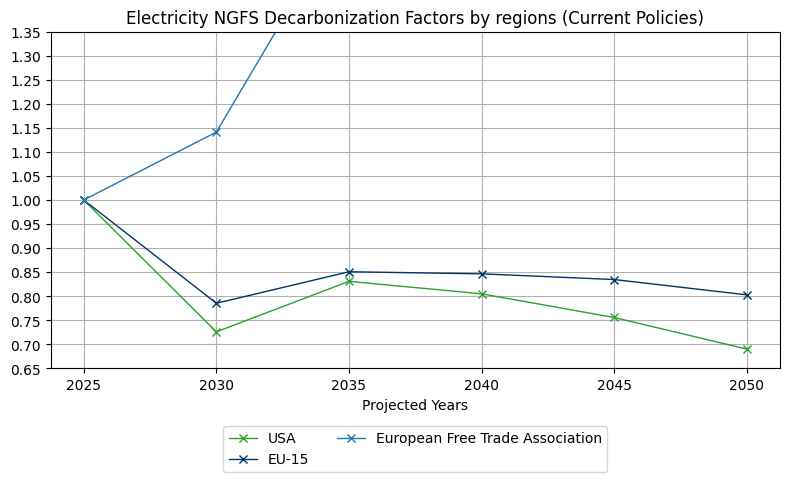}\end{minipage} 
\caption{Regional Electricity decarbonization trajectories} 
\label{fig:electricity_regions} 
\end{figure}

A stronger decarbonization effect is observed for MSCI USA. To explain this difference, Figure~\ref{fig:electricity_regions} compares the regional \textit{Electricity} trajectories entered in the calculation.  

The results highlight that the US \textit{Electricity} trajectory declines more rapidly than the corresponding EU-15 trajectory. Because MSCI USA is uniquely exposed to the US pathway, whereas the Bloomberg Euro Corporate Bond incorporates a mix of US and European higher exposures, the resulting \textit{Electricity} decarbonization factor is stronger for MSCI USA. This regional effect explains most of the difference observed for the \textit{Utilities} sector\footnote{The EU FTA decarbonization factor trajectory increases quite rapidly due to emissions close to zero in this region. This behavior will be very limited on \textit{Electricity} pathways for Bloomberg Euro Corporate Bond since there is little market capitalization weight in EU FTA.}.

\paragraph{Synthesis} The divergence between the normalized emissions pathways of MSCI USA and the Bloomberg Euro Corporate Bond can be attributed to two distinct mechanisms. First, a sector-composition effect arises from the presence of the \textit{Materials} sector in MSCI USA and its absence from the Bloomberg Euro Corporate Bond. Given its substantial emissions weight and strong decarbonization factor, this sector contributes significantly to the faster decline of the MSCI USA pathway. Second, a geographical effect emerges within the \textit{Utilities} sector. Although the sector carries comparable emissions weights in both ETFs, regional differences in NGFS sector \textit{Electricity} trajectories lead to a stronger decarbonization factor for MSCI USA. This effect is primarily driven by the more rapid decline of the US \textit{Electricity} pathway relative to its European counterparts. Together, these two effects explain most of the divergence observed between the normalized emission pathways of the two ETFs.

%% file: 3c_Discussions_conclusions.tex
\subsection{Key Takeaways and limits} 
\paragraph{Key Takeaways} This case study highlights several important insights regarding the proposed sectoral decarbonization approach.
It provides a transparent framework for assessing how portfolio emissions may evolve under different climate transition scenarios. In this paper, the GICS-NGFS mapping allows a usage of public climate scenarios from NGFS and translate their impacts on portfolios emissions pathways at a GICS sectors level. The practical study focused the analysis on the NGFS Current Policies scenario but it can be easily led on any GCAM pathway. In general, this approach can be extended to any climate scenario, as long as a GHG mapping translating the sectors between scenarios and portfolios is provided. 

The methodology enables a complete decomposition of portfolio-level results into sectoral and regional contributions, improving the interpretability of emissions pathways, as shown in the previous applications. 

The framework can be applied consistently to any asset type or index as long as their geographical and GHG emissions decompositions are known. 

\paragraph{Limits} One key limitation of this proposed approach is the exclusion of Scope 3 emissions. Although they often represent a substantial share of the overall carbon footprint of companies and are therefore highly relevant for transition risk assessments, their projection remains particularly challenging due to limited data availability, heterogeneous reporting practices, and significant methodological questions. Consequently, this paper focused on Scope 1 and Scope 2 emissions, for which disclosures are generally more robust and comparable across firms. Future research could extend the proposed framework to incorporate Scope 3 emissions as data quality and scenario methodologies continue to improve.

Another limitation of the analysis lies in the assumption that the portfolio composition remains static throughout the projection horizon. In practice, portfolio holdings evolve continuously as investors adjust their allocations in response to market conditions, regulatory changes, climate-related considerations... Keeping fixed portfolio weights eases the interpretation of the emissions pathways, as it focuses only on the effect of sectoral and geographical decarbonization pathways. However, this assumption may not accurately reflect the emissions trajectories of real-world portfolios over the long term. Indeed, portfolio decarbonization may result not only from reductions in companies' emissions intensities, but also from shifts in sectoral, regional, or issuer exposures. Future research could therefore extend the framework by incorporating dynamic portfolio allocation assumptions and investment strategies under climate transition scenarios.

Finally, the projected emissions trajectories depend on the assumptions embedded in the selected climate scenarios. Sectoral decarbonization factors reflect complex macroeconomic, technological, and policy developments that may be difficult to separate and interpret individually. Consequently, portfolio managers should exercise caution when using these projections for climate stress testing and avoid interpreting the resulting emissions pathways as forecasts.

%% file: 4_Conclusions.tex
This paper proposes a NGFS-GICS mapping framework designed to translate climate-scenario emissions pathways into a classification system widely used by financial market participants. By explicitly incorporating the materiality of Scope 1 and Scope 2 GHG emissions, the proposed approach provides a practical bridge between NGFS climate scenarios and portfolio-level climate analysis. Combined with the proposed sectoral decarbonization approach, the methodology enables the projection of financed-emissions pathways while preserving the sectoral structure commonly used by asset managers and financial institutions. The application to two illustrative ETFs demonstrates that the framework not only produces portfolio emissions trajectories under alternative climate scenarios but also identifies the sectoral and geographical drivers underlying these trajectories. As such, the approach offers a transparent and operational tool for climate alignment assessments, transition risk analysis and long-term portfolio monitoring. Future research could extend the framework to Scope 3 emissions, alternative sectoral classifications and dynamic portfolio allocation assumptions, further strengthening its usefulness for climate-related financial decision-making.

%% file: 5_Acknowledgments.tex
\noindent The authors thank Lina JOUKER, Nestor TOROMAN, Philomène CHOLET, and Juliette MARTY for contributing to the methodology development at its different stages, providing their useful technical expertise. Additionally, we thank EDF Lab and the FiME laboratory\footnote{https://fime-lab.org/}. Finally, we specially thank Malak AL KOSTIT, for her work on the tool implementation.

Research work realized thanks to the Pladifes COGEM project in the context of the Pladifes EquipEx (ANR-21-ESRE-0036), hosted at the Institut Louis Bachelier.